**MindPortal**

# MindSpeech: Continuous Imagined Speech Decoding using High-Density fNIRS and Prompt Tuning for Advanced Human-AI Interaction

Suyi Zhang[a,] Ekram Alam[a],Jack Baber[a], Francesca Bianco[a], Edward Turner[a], Maysam Chamanzar[a,b,c,d], Hamid Dehghani[a,e]

[a] *MindPortal Ltd, London, EC2A 3PG, United Kingdom*
[b] *Department of Electrical and Computer Engineering, Carnegie Mellon University, Pittsburgh, PA 15213, USA*
[c] *Department of Biomedical Engineering, Carnegie Mellon University, Pittsburgh, PA 15213, USA*
[d] *Neuroscience Institute, Carnegie Mellon University, Pittsburgh, PA 15213, USA*
[e] *School of Computer Science, University of Birmingham, Birmingham B15 2TT, United Kingdom*

| **Dr Suyi Zhang** | **Dr Francesca Bianco** | **Edward Turner** | **Ekram Alam** | **Jack Baber** |
|---|---|---|---|---|
| Senior Research Engineer | Senior Research Scientist | Research Engineer | CEO & Concept Architect | CTO & Technical Architect |

| **Prof Maysam Chamanzar** | **Prof Hamid Deghani** |
|---|---|
| Scientific Advisor | Scientific Advisor |

**Abstract**: In the coming decade, artificial intelligence systems will continue to improve and revolutionise every industry and facet of human life. Designing effective, seamless and symbiotic communication paradigms between humans and AI agents is increasingly important. This paper reports a novel method for human-AI interaction by developing a direct brain-AI interface. We discuss a novel AI model, called MindSpeech, which enables open-vocabulary, continuous decoding for imagined speech.

This study focuses on enhancing human-AI communication by utilising high-density functional near-infrared spectroscopy (fNIRS) data to develop an AI model capable of decoding imagined speech non-invasively. We discuss a new word cloud paradigm for data collection, improving the quality and variety of imagined sentences generated by participants and covering a broad semantic space. Utilising a prompt tuning-based approach, we employed the Llama2 large language model (LLM) for text generation guided by brain signals. Our results show significant improvements in key metrics, such as BLEU-1 and BERT P scores, for three out of four participants, demonstrating the method's effectiveness.

**MindPortal**

Additionally, we demonstrate that combining data from multiple participants enhances the decoder performance, with statistically significant improvements in BERT scores for two participants.

Furthermore, we demonstrated significantly above-chance decoding accuracy for imagined speech versus resting conditions and the identified activated brain regions during imagined speech tasks in our study are consistent with the previous studies on brain regions involved in speech encoding. This study underscores the feasibility of continuous imagined speech decoding. By integrating high-density fNIRS with advanced AI techniques, we highlight the potential for non-invasive, accurate communication systems with AI in the near future.

# 1. Introduction

An AI model which can effectively translate the thoughts a human is imagining into text or other mediums would revolutionise human-AI communication and communication between humans and computers in general.

To produce such an AI model, brain signals of humans imagining thoughts, words and sentences in their mind need to be recorded to supply the AI with the appropriate data to decode. To do this, hardware sensors which can monitor brain activity with high spatiotemporal resolution need to be deployed. Surgically implanted electrodes have shown some potential for speech decoding (Brumberg et al., 2016), but they are invasive. Non-invasive techniques such as fMRI have great potential for decoding neural signals during reading and listening tasks (Huth et al., 2016; Tang et al., 2023; Ye et al., 2024). However, fMRI requires bulky, expensive equipment, which makes this neural recording technology non-portable and more importantly, not under naturalistic conditions. In this work, MindPortal opted to use functional near-infrared spectroscopy (fNIRS) for functional neural imaging due to its portability and cost-effectiveness, monitoring brain oxygenation changes with performance similar to fMRI (Cao et al., 2018; Eggebrecht et al., 2012).

There have only been limited recent studies on using fNIRS to study and decode imagined speech such as Cao et al. (2018) and Rybář et al. (2021) which have explored decoding of semantic content from brain signals. Additionally, Naseer and Hong (2014) applied fNIRS to develop an online binary decision decoding system for responses like "yes" and "no." MindPortal's previous work, titled 'MindGPT' (Zhang et al., 2024) also demonstrated that an AI model can classify semantically different sentences leveraging fNIRS for neural recording during imagined speech. While these studies highlight fNIRS's potential for semantic decoding, they primarily focus on single-word decoding or sentence classification, limiting their application to more extensive segments of imagined speech. To date, developing an open vocabulary, continuous decoder for imagined speech has remained elusive, and using the portable method of high density fNIRS takes the state of the art one step further towards achieving this goal.

Recently, large language models (LLMs) have dominated computational language modelling, generating coherent language from text prompts. However, traditional methods separate brain decoding and language generation into two phases, limiting their integration. A recent paper introduced BrainLLM (Ye et al., 2024), a novel approach based on LLM prompt tuning that directly incorporates semantic representations from non-invasive fMRI recordings into the language generation phase. This method eliminates the need for post-hoc candidate selection, enhancing the generation of open vocabulary continuous speech directly from brain signals, outperforming previous models reliant solely on text prompts. This study used fMRI data only from reading and listening tasks, and was not tested on imagined speech data or from other more portable non-invasive neuroimaging methods.

Our study aims to develop an open vocabulary, continuous decoder for imagined speech high-density fNIRS data. We first introduce a novel paradigm for imagined speech fNIRS data collection, that captures a wide variety of semantic meanings. The novel paradigm prompts participants to imagine sentences of different topics by providing topic words and some additional keywords during the task. The participants are also asked to type out the imagined sentences as ground truth after the task. Differently to our previous research with MindGPT (Zhang et al., 2024), the choice of this free-form paradigm allows us to isolate imagined speech from other cognitive functions, e.g., memory, by eliminating the reliance on memorisation of preset sentences and minimises the need for training to collect imagined speech data. Secondly, a finite impulse response (FIR) model was used to systematically identify the appropriate BOLD delay for imagined speech fNIRS data. Finally, we adapted the prompt tuning-based BrainLLM model for natural text generation to decode imagined speech fNIRS brain signals into text.

We demonstrated that using a prompt tuning approach with brain signal-guided inputs results in the generation of sentences that significantly outperform those created from randomly paired brain signals and context inputs. These generated sentences show higher scores in both language and semantic similarity metrics, indicating the effectiveness of our method.

# 2. Material and Methods

## 2.1 Participants

This study included a total of 4 participants (male = 3, female = 1), ranging in age from 25 to 28 years, with a mean age of 26.5 years. Participants were required to be fluent in English, free from any known neurological disorders, and not currently undergoing any form of psychological or psychiatric treatment. Participants were also asked to refrain from drinking coffee and make use of any other substance that may alter their mental state before each data collection session. Participants wearing glasses were asked to use contact lenses for data collection for ease when setting up the fNIRS neuroimaging cap. All participants provided informed consent and were compensated for their time. All participants completed the tasks in less than 16 sessions.

## 2.2 Experimental Paradigm

Inspired by the related word presentation in Experiment 1 in Pereira et al., 2018, we created a task specifically for continuous imagined speech generation called a 'word cloud' task (Figure 1). Our word cloud task consisted of participants being presented with a selection of words on the screen. Differently to Pereira et al., 2018, participants needed to imagine a sentence from the given words sequentially. The central word in a larger font was the topic word, and the words in a smaller font surrounding the topic word were the keywords related to the topic. During the trial, 3 keywords were selected randomly, one at a time, by providing the word in bold font, each for a period of 7s. As the keyword appeared in the bold font, participants were asked to imagine a sentence relating both the topic word and one of the keywords. Participants were required to use only the last bolded keyword and the topic word not all previously highlighted keywords. After the presentation of all the keywords, the task was completed for the given topic and participants were instructed to type the sentences they imagined relating the topic to each keyword, using a computer keyboard. This was used as a ground truth for the decoder training. The typing period was not timed, and participants were prompted to move on to the next topic after they had completed all typing. The word cloud task aims to tap into participants' imagined speech cognitive function and is distinctly different from silent reading, where participants are asked to read sentences from a screen. In addition, as opposed to our previous work on MindGPT (Zhang et al, 2024), there is no heavy reliance on memorisation in this word cloud task or need for pacing imagined speech words at a specific word/minute rate using a metronome, removing any contamination of other cognitive functions, such as memory or auditory processing, from our paradigm to isolate imagined speech, as well as providing a new way to prompt for imagined speech in participants without much prior training.

A

B

*Figure 1. Illustrations of the word cloud paradigm. A. A participant imagines a sentence given a topic word and a highlighted keyword in the 'word cloud' paradigm. After the imagine period concludes for the topic, they are asked to type out the imagined sentences. The brain signals during imagined speech and the ground truth texts are used for decoder training. B. An example topic block shown to the participant during the experiment. The topic word is the*

*bolded word with larger font in the middle, and the surrounding keywords are selectively highlighted in random order to prompt the participant to imagine a sentence related to both the topic and the keyword. The participant is then asked to type out the sentences imagined in the block at the end.*

In this study, a total of 272 distinct topics were included, of which 147 topics and related keywords were taken from the reading tasks in Pereira et al., 2018 (Experiments 2 and 3). This selection of topics covers a wide range of concepts in the form of concise, informative sentences as presented in Pereira et al., (2018). The additional 125 topics were generated using keywords from the New General Service List (NGSL) Spoken (Browne, 2014), which covers over 90% of the most frequently used words in the English language. A total of 3 keywords were grouped under each new topic not used in Pereira's topic sets. OpenAI's GPT-4 model (2024) was then used to generate sentences of similar lengths to those used in the Pereira dataset, and then additional keywords were extracted from these sentences for display during the task. These GPT-4 generated sentences and keywords were generally less formal, reflecting the verbal nature of the NGSL Spoken data, thereby balancing the full dataset in terms of verbal styles, for a more complete representation of the English language.

A word cloud task of three topics formed a run. Each run contained a resting trial of 20s, which was randomly added at the beginning, middle or end of each run. During resting trials, the word 'baseline' appears on the screen throughout the trial. A topic break of 10s was used if no resting trial was added in between topics. On each day of the experiment, participants were asked to complete as many runs as possible in an hour, before removing the cap for a 20 minute break (or longer if required). All the runs within an experimental day were grouped into a session. All 272 topics were used for the word cloud task in a randomised order. However, due to the limited amount of experimental time, participants were able to complete only a portion of the topics (see Table 1 for topic and sentence statistics). Note that only brain signals obtained during the imagined period (i.e., keyword highlighting) were used in the decoder training, not the typing period, to minimise the movement artefacts.

## 2.3 fNIRS Setup and Data Preprocessing

In this experiment, we used a commercially-available Continuous-Wave (CW) high-density 48x48 fNIRS system (NIRx Inc.) for collecting neurovascular data. The montage provided full-head coverage, including 48 sources and 47 detectors (the extra detector is used for the short-distance channels) (Figure 2). The system consists of a total of 388 channels (194 with a light source wavelength of 760 nm and 194 with a light source wavelength of 850 nm), with channel distances ranging from ~21 mm to ~42 mm, and 8 short-distance channels (channel distances: < 10 mm). A sampling rate of 5.9Hz was used. Further details of the experimental setup can be found in our previous work, MindGPT (Zhang et al., 2024).

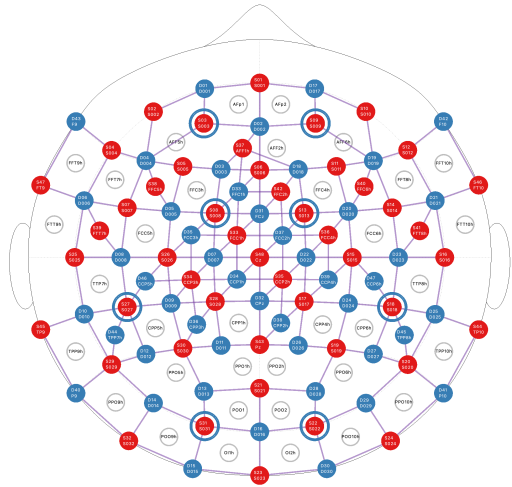

*Figure 2. Whole head high-density fNIRS montage, including 48 sources, 47 detectors and 8 short-distance channels.*

Brain data collected during the experiment was streamed through the NIRx acquisition software, Aurora fNIRs (NIRx Medical Technologies LLC), and saved in XDF format. The saved data was then loaded and subjected to a series of preprocessing techniques. For each run, data underwent conversion of raw signals to optical density, detrending, short channel regression correction, motion artefact removal, conversion to haemoglobin concentration using a partial pathlength factor (ppf) of 6, bandpass filtering between 0.01 and 0.7 Hz (Huppert et al., 2009). For an imagined sentence recorded in 7s, the corresponding data has a sequence length of 42 time points by 388 channels (194 oxy-haemoglobin and 194 deoxy-haemoglobin channels).

## 2.4 Finite Impulse Response (FIR) model

A Finite Impulse Response (FIR) model (Dale, 1999) was employed to estimate the brain's hemodynamic response to different stimuli without assuming a specific shape for the response function. Given that the brain's haemodynamic response may vary across individual participants and experimental tasks, a FIR model was estimated for each participant, consisting of two conditions of word cloud and rest. The raw intensity data was converted to optical density and then to haemoglobin concentrations and resampled to 0.5Hz. The FIR model uses a design matrix with columns representing different time lags of the stimulus, allowing the response at each time point following the stimulus to be captured. In a first level design matrix, the hemodynamic response function (HRF) was set to "fir" with 5 delays. A cosine drift model with a high-pass filter set at 0.01 Hz was applied to account for low-frequency drifts. The design matrix was created without oversampling. Averaged short channel sequences were used as nuisance regressors in a General Linear Model (GLM) to isolate brain activity from systemic physiology. Individual design matrices were created and estimations were performed for each run, followed by the concatenation of all the results across the runs for each participant. Finally, mixed-effect linear models were used to estimate coefficients separately for each delay, task, and oxy/deoxygenated haemoglobin.

## 2.5 Continuous Imagined Speech Decoding Model

To decode continuous imagined speech from non-invasive fNIRS brain signals, the decoding model needs to be able to a) extract semantic information from the brain signals, and b) generate legible continuous speech given the semantic information. A recent paper described a novel approach that combines both processes in their BrainLLM model (Ye et al., 2024), which demonstrated satisfactory decoding performances in open fMRI datasets with reading and listening paradigms. Inspired by this model, we adapted the processes and used it to decode imagined speech from our fNIRs data.

### 2.5.1 Prompt tuning for foundational LLMs

Tuning-based methodologies have been previously employed to harness the extensive knowledge embedded within large language models (LLMs) for decoding time-series data (Jin et al., 2024, Liu et al., 2023). These approaches generally encompass the processes of segmenting and tokenizing time series signals and associated textual data, followed by fine-tuning the models for specific tasks.

The prompt tuning process comprised of the following steps:

1. Brain recordings were collected during the imagined speech condition, during which participants silently imagined a sentence based on a given topic word and a keyword ('Word Cloud' task, see section 2.2 and Figure 1).
2. The ground truth sentence, which participants imagined and subsequently typed from memory using a keyboard, was segmented into patches (Figure 3A). The initial segment served as the *context input*, while the subsequent segment served as the *continuation*. Correspondingly, the fNIRS brain recordings associated with the continuation segment were extracted. A 6s delay was applied to the fNIRS data, as identified by the FIR model (see section 2.4).
3. The brain signals underwent processing through a brain decoding model that mapped the time series data to standard LLM embeddings. Simultaneously, the context input text was tokenized and converted into LLM embeddings.
4. The context input embeddings were concatenated with the brain signal-generated embeddings, forming the prompt input to the LLM (Figure 3B). During training, the brain decoding model was trained to learn the mapping from brain signals to LLM embeddings, while the LLM had its parameters frozen during this process.
5. The LLM generated the continuation text based on the concatenated prompt input. The predicted continuation was then compared with the ground truth continuation text to facilitate the training of the encoding model.

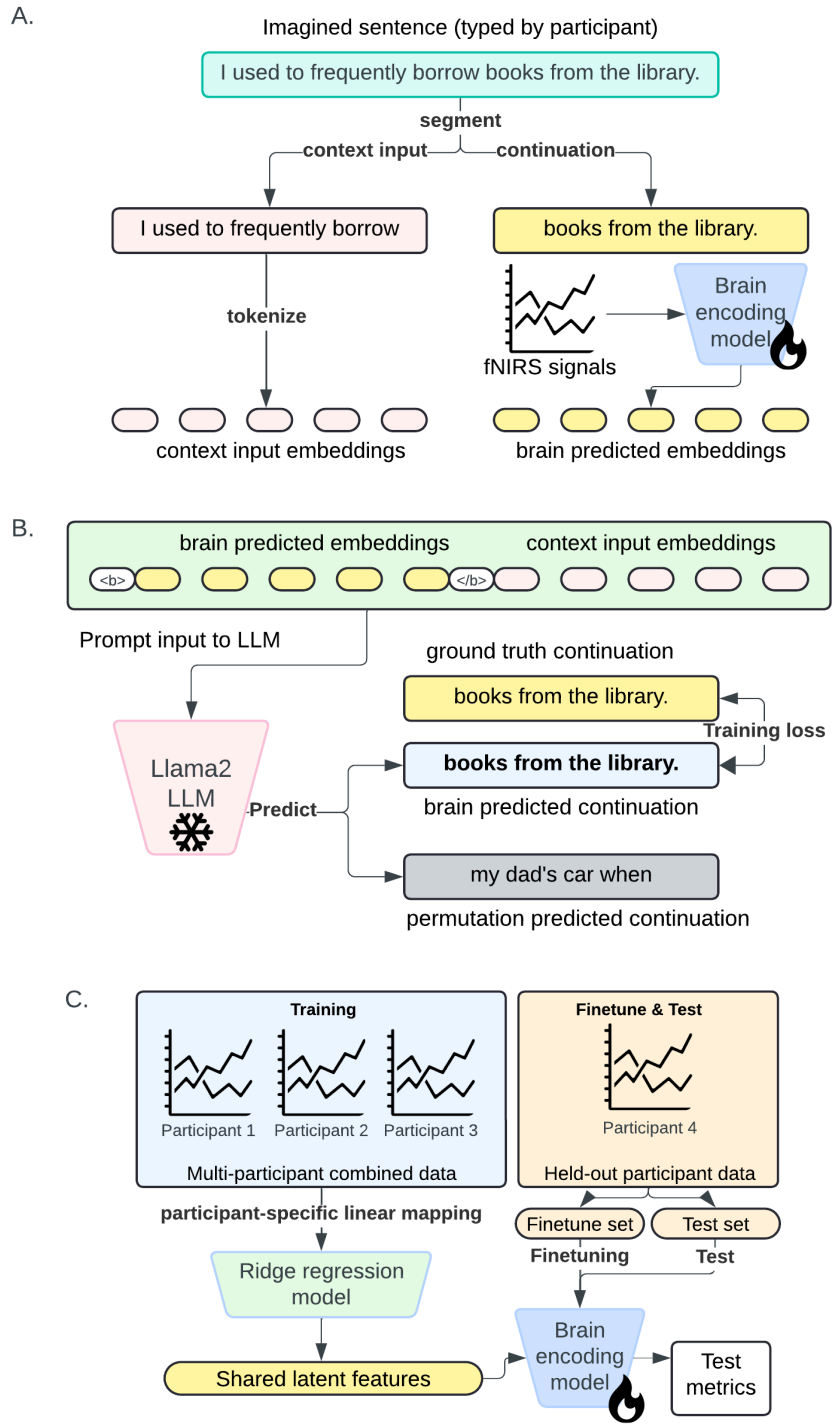

*Figure 3. Block diagrams showing the prompt tuning process in continuous imagined speech decoding. A. The ground truth sentence, which the participant imagined and subsequently typed from memory, is segmented into context input and continuation. LLM embeddings are generated from the context input text, and the fNIRS brain recordings associated with the continuation segment goes through a brain encoding model to generate brain predicted LLM embeddings. The fire symbol represents the weights in the brain encoding model that are trainable. B. The brain signal-generated embeddings are concatenated with the context input embeddings, forming the prompt input to the Llama2-7b LLM with frozen parameters (snow*

*symbol). The predicted embeddings are converted to text and compared with the ground truth continuation text in order to generate a training loss. Permutation predictions are generated with permutation inputs, where brain signals corresponding to one sentence are paired with the context input from another sentence randomly chosen in the same participant. C. Brain data from multiple participants, except the held-out test participant, are aligned using a ridge regression model to create a shared latent feature space, before being inputted to the brain encoding model training. The model is then fine-tuned with 100 trials and metrics are evaluated on 200 held out test trials from the test participant .*

## 2.5.2 Data preparation

In our study, Llama2-7b (Touvron et al., 2023, [https://huggingface.co/meta-llama/Llama-2-7b-hf](https://huggingface.co/meta-llama/Llama-2-7b-hf)) was used as the LLM for text processing and generation. An LLM works by leveraging a deep neural network, typically with billions of parameters, that has been trained on vast amounts of text data to predict and generate human-like text based on input text prompts. It uses this training to understand context, grammar, and semantics, allowing it to generate coherent and contextually-relevant responses.

Following Ye et al., 2024, each ground truth sentence imagined by the participant was divided into segments. Given that the imagined sentences were in general shorter than the original Wikipedia-style reading sentences in Pereira et al., 2018, each sentence was segmented into two parts only (context input and continuation, see Figure 3A). Note that any typed sentence with three words or fewer and their related fNIRS data were excluded from the continuous imagined speech decoder training, as they were considered too short for segmentation. The texts were then tokenized into token IDs with a maximum length of 32 with padding. These token IDs were then converted to Llama2 embeddings afterwards. Additional custom tokens <brain/> and </brain> were added to the tokenizer, which were used to separate brain generated embeddings from the context input embeddings. The sample numbers of the brain signals corresponding to the continuation part of the sentence were extracted from the preprocessed fNIRS signals. A fixed delay of 6s was added to the brain data in order to account for BOLD delays (see FIR model section for delay estimation).

For permutation inputs, the continuation brain signals from a given sentence were paired with the context input text from another sentence randomly chosen from the participant. This method kept the distribution of the participant's brain signals compared to using random numbers or scrambled brain data. Note that permutation *does not* refer to the permutation of the time points or channels of the brain signals, but instead the permutation of the sentences.

Additional experimental conditions of context text only and brain signal only were also included. Context only refers to where the LLM uses only context input text for generating continuation predictions, and brain signal only refers to where brain signals alone are used for continuation predictions. These conditions followed the same data processing as described for brain and permutation input conditions mentioned above.

For all experimental conditions, the LLM embeddings for context input and continuation brain signals for all trials within a participant were stored in a Python pickle file, and were loaded as Pytorch datasets during training. The brain data was z-scored within each session for each participant in order to minimise the effects of different cap placement across different days of experiments.

## 2.5.3 Brain encoding model

The brain encoding model maps fNIRS data to LLM embeddings. It takes a time-series input of shape [sequence length x number of channels], where sequence length is the number of time points in the brain signals segment corresponding to the continuation part of the sentence, and number of channels is the total number of fNIRS channels of 388.

A sequence-to-sequence (Seq2Seq) neural network model with transformers was used as the brain encoding model (Sutskever et al., 2014, Vaswani et al., 2017). The Seq2Seq model is a versatile neural network capable of handling variable-length sequences effectively, compared to other simpler deep neural networks. For this study, a transformer model was used for both its encoder and decoder components. Transformers utilise a self-attention mechanism, which allows the model to weigh the importance of different time steps dynamically, regardless of their distance in the sequence. This capability is particularly advantageous for time series data such as fNIRS data, where important patterns and dependencies can occur over various time scales.

A transformer with a single encoder layer and a fully connected layer was used, with a fixed dropout rate of 0.3. As an encoder, it processes the input sequences of size 388 (number of fNIRS channels), with hidden size of 100. As a decoder, it takes the input sequence of size 100 from the encoder, and outputs sequences of shape 4096 (Llama2 embedding size). During the forward pass, the source sequence was first processed by the encoder, and the encoder's output was subsequently passed to the decoder to generate the final output sequence. The weights of the transformer model were initialised using Xavier uniform distribution for projection and linear weights, while setting biases to zero or a small constant, and applying similar initialization to the linear layers.

## 2.5.4 Model training, validation and testing

The training process involved two stages, a pretraining step and a main training step. Pretraining was run for 10 epochs before the main training started, with an initial learning rate of $1e^{-3}$. The pretraining loss computes the mean squared error (MSE) loss between the brain encoding model's predicted embeddings and the mean value of the ground truth continuation text embeddings. The ground truth texts were converted into embeddings, which were then fused into a mean representation to ensure a constant length. This approach also minimises the probability of information leak from the ground truth text.

During main training, the parameters of the Llama2 LLM were frozen, preserving its inherent knowledge while utilising its ability to generate coherent text outputs given limited input examples (Liu et al., 2023). Instead, the brain encoding model was being trained in an end-to-end pipeline in order to learn to generate LLM embeddings from brain signals. The training function initialised data loaders, where 200 trials were kept out for testing, the rest of the dataset were divided into 80% training and 20% validation. Training batch size was set at 8. Adam optimizer (Kingma et al., 2017) with an initial learning rate of $1e^{-4}$ and a learning rate scheduler with step=1 was used for training. The model weights are saved if a lower validation loss is achieved in the next epoch, and early stopping is implemented if the validation loss does not improve over 10 epochs. The max number of epochs was set to 50.

Training loss in main training was defined as the cross-entropy loss between the Llama2 predicted token logits and the true labels in the continuation text. LLM's predicted token logits were the raw, unnormalized scores output by the LLM for each token in the vocabulary, indicating the likelihood of each token being the next in the sequence, before applying a softmax function to convert them into probabilities. The predicted and true sequences were aligned in lengths, and a mask extracted the portion of the predicted sequence corresponding to the true labels. The loss was then calculated using the cross-entropy function on the filtered logits and true labels, with the `reduction` parameter set to "mean" to average the loss over the batch. This process ensures that the loss computation only considers the valid parts of the generated sequence, ignoring padding or other irrelevant tokens. In each training epoch, the gradients were reset to zero, the loss was backpropagated, and gradient norms were clipped to a maximum of 10 before updating the model parameters using the optimizer, with the total loss being accumulated for each batch.

Finally, the held out test trials were used for text generation and metric calculations. Predicted text tokens were generated given the concatenated prompt input from the context input and the brain predicted embeddings. The tokens were converted back to text using Llama2's tokenizer, where special tokens such as </s> and <s> are removed. The LLM predicted text and ground truth text for the continuation part of the sentence are then used as prediction and hypotheses for a series of metric calculations.

## 2.5.5 Multi-participant alignment and fine tuning

We additionally tried out a method by combining multiple participant's data for training in order to maximise data usage, as well as increasing coverage of the overall semantic map (Figure 3C). Data from multiple participants except for the left out test participant was first aligned for training. This was accomplished by passing the brain data to a ridge regression model that maps the input from 388 channels to 100 features. The ridge regression model consists of multiple linear layers, and processes data based on the given participant ID, so that each participant effectively has a different feature extractor and the resulting features reside in a latent feature space shared by all the training participants. This process aims to extract features and reduce noise from fNIRS signals. The extracted features are used as input to the brain encoding model instead of the full fNIRS signals.

A leave-one-participant-out test was used, where one participant's data was left out and all other participants' data was used for training. Before testing metrics were calculated, the pre-trained brain encoding model from multiple participants was first finetuned with 100 fine-tune trials from the test participants' data. The data collection time for 1 trial is about 22s (7s for the imagined period, 15s for typing based on a typical typing speed of 40 words/min and the averaged sentence length of 10 words), which means the collection of 100 trials for fine tuning would have taken about 37 minutes to complete if collected in a separate run. The 200 test trials were held out from the finetuning process. The previously saved cache model was loaded and trained with the same procedures and parameters as mentioned in 2.5.4 with the finetune trials, then the test metrics are calculated on the held out test trials.

### 2.5.6 Evaluation metrics

Natural language processing metrics are used to evaluate model performance by comparing the model generated sentences with the ground truth sentences which the participant was imagining. These include BERTScores, a text evaluation metric that assesses the similarity between candidate and reference texts by computing the cosine similarity of their contextual embeddings derived from the BERT model (Zhang et al., 2020). This is further divided into 3 scores - 1) BERT F1, which combines precision and recall to measure accuracy by balancing both false positives and false negatives; 2) BERT P (Precision), a precision metric focusing on the correctness of the positive predictions made by the model; and 3) BERT R (Recall), a recall metric assessing the model's ability to capture all relevant instances. 4) BLEU-1 score, which measures the overlap of n-grams (unigrams) between the generated sentence and the reference, indicating surface similarity (Papineni et al., 2002). 5) METEOR scores, which considers precision, recall, and synonymy, providing a more nuanced similarity score than BLEU (Banerjee & Lavie, 2005). 6) ROUGE L score focuses on the longest common subsequence between the generated and reference sentences, reflecting structural similarity (Lin, 2004). 7) Word Error Rate (WER), which calculates the edit distance between the generated sentence and the reference, indicating accuracy (Klakow & Peters, 2002). These multiple metrics provide a comprehensive evaluation of different aspects of model performance.

## 2.6 Imagined Speech Detection Decoder

In our previous work with MindGPT (Zhang et al., 2024), we showed that an Extra Trees Classifier (XTC) model was able to differentiate imagined speech from the rest condition with an average accuracy of 66% across four participants (best average accuracy: 71%). Our previous imagined speech paradigm however relied heavily on memory and participant training before data collection. We sought to repeat this test with this new fNIRS dataset of imagined speech data from four participants collected while performing the word cloud paradigm, to show that imagined speech detection using fNIRS is possible without relying on other cognitive functions or need for pacing imagined speech words at a specific word/minute rate using a metronome, removing any contamination of other cognitive functions, such as memory or auditory processing. Therefore, we trained a new XTC model to decode imagined speech from the rest data in fully preprocessed brain signals for each participant from this study. The same parameters and methods as what we used in our previous MindGPT (Zhang et al., 2024) work were used. All models were trained on each participant to determine the best individual accuracy, although results from participants were also averaged to identify overall performance of our models in successfully classifying imagined speech vs rest condition.

## 2.7 Imagined speech related brain activations

In addition to applying decoding models to the imagined speech data, analyses involving haemodynamic response modelling and GLM-based statistical testing were conducted in order to show imagined speech related activations in the brain. To extend our knowledge of semantic representation in the brain, brain activations from the contrast 'word cloud > rest' were compared across participants.

The fNIRS data was preprocessed to haemo data using the steps described in section 2.3. A first-level design matrix was constructed to model the hemodynamic response associated with neural activity, using the python package mne-nirs (version 0.6.0). Channels (distance > 10mm) are used in the general linear model (GLM, Friston et al., 1994), with a cosine function to model and correct for low-frequency drift in the signal, and a high-pass filter of 0.005Hz to remove slow signal variations not contributed to neuronal activities. The hemodynamic response function (HRF) model adopted was based on the Statistical Parametric Mapping (SPM) approach, a standard model for estimating the brain's vascular response to neural activity. Lastly, a stimulus duration of 7s from the onset of each word cloud sentence (i.e. keyword highlight) and each resting period was considered. Averaged short channel sequences (distance < 10mm) were included as nuisance regressors in the design matrix. Conditions 'word cloud' and 'rest' were specified and the GML parameters were estimated. The contrast 'word cloud > rest' was then estimated from GLM theta values, and z-scores are calculated. Surface plots are generated with the estimated z-scores. Due to computing memory constraints, only 10 runs were used for each participant, totalling to around 50000 time points per participant. For group analysis, a group-level mixed linear model was run using the statsmodels package, fitting data with participant IDs as grouping, and the model was optimised using the "nm" method.

To verify the accuracy of our method to localise brain areas recruited during the cognitive tasks and compare this accuracy in localisation across participants, we also conducted finger tapping experiments at the start and end of each session and compared brain activation in response to right vs left finger tapping and their localisation within the cortex. Right vs left finger tapping tasks have been previously shown to lead to a quite strong and clear differential activation in the contralateral motor cortex, with respect to the hand completing the finger tapping (e.g., Batula et al., 2017). Therefore, identifying the correct localisation of brain activation during the finger tapping experiment would confer us with increased confidence with respect to precise cap placement and brain region coverage, as well as good quality of data collected. See *Supplementary Figure S1* for a brief explanation of the finger tapping paradigm used, as well as the brain activations.

# 3. Results

## 3.1 Dataset statistics

Statistics of each participant's total number of topics and sentences completed for the word cloud tasks are summarised in Table 1. Participants completed on average 216 topics in the word cloud task, although the number of topics varied across participants. The average number of words per sentence also varied across participants, where participant 4 had on average the longest imagined sentence length of 10.5 words, while participant 3 had the shortest averaged sentence lengths of 8.67 words. However, participant 1 used the most unique words per sentence of 3.28 words compared to other participants. Summarising language styles using statistics is challenging; however, participants exhibited varied writing styles in addition to differences in sentence lengths and unique word choices. The imagined sentences exhibit a more conversational and informal tone (see Table 3 for examples), characteristic of spoken language, rather than the structured and formal style typical of Wikipedia-style text in Pereira et al,. (2018).

*Table 1. Statistics on total number of sentences, topics and keywords imagined by the participants during the data collection*

| Subject | Total sentences | Unique topics | Unique keywords | Total unique words | Avg. sentence length | Avg. unique words per sentence |
|---|---|---|---|---|---|---|
| Participant 1 | 433 | 111 | 396 | 1420 | 9.26 | 3.28 |
| Participant 2 | 619 | 216 | 591 | 1803 | 8.92 | 2.91 |
| Participant 3 | 767 | 269 | 732 | 2015 | 8.67 | 2.63 |
| Participant 4 | 827 | 267 | 737 | 2416 | 10.52 | 2.92 |
| Mean | 661.5 | 215.75 | 614 | 1913.5 | 9.34 | 2.94 |
| Std | 175.63 | 74.02 | 160.32 | 415.76 | 0.82 | 0.27 |

## 3.2 Identifying BOLD delays with FIR models

To determine the optimal BOLD delay to use in the imagined speech decoding model training, we employed a FIR model to estimate coefficients across a range of delays. FIR models do not assume a fixed haemodynamic response function, allowing for the estimation of individual haemodynamic responses and accommodating inter-individual differences. The delay with the highest coefficient from these estimations is considered the optimal delay. Figure 4 presents the delay coefficients from a group estimation for all runs within each participant, categorised by word cloud and resting conditions, as well as oxy/deoxygenated haemoglobin (HbO/HbR). For participants 1, 2, and 3, the highest coefficient for HbO occurred at the third delay (equivalent to a delay of 4-6 seconds, given a resampling rate of 0.5 Hz), while for participant 4, it was at the fourth delay (6-8 seconds). The second highest delay for participants 1 and 3 was 6-8 seconds. Based on these findings, a delay of 6 seconds is applied to the fNIRS data to accurately account for the BOLD responses in word cloud tasks. This standardised delay was chosen to maintain consistency across participants and simplify the model implementation, ensuring reliable comparison of results.

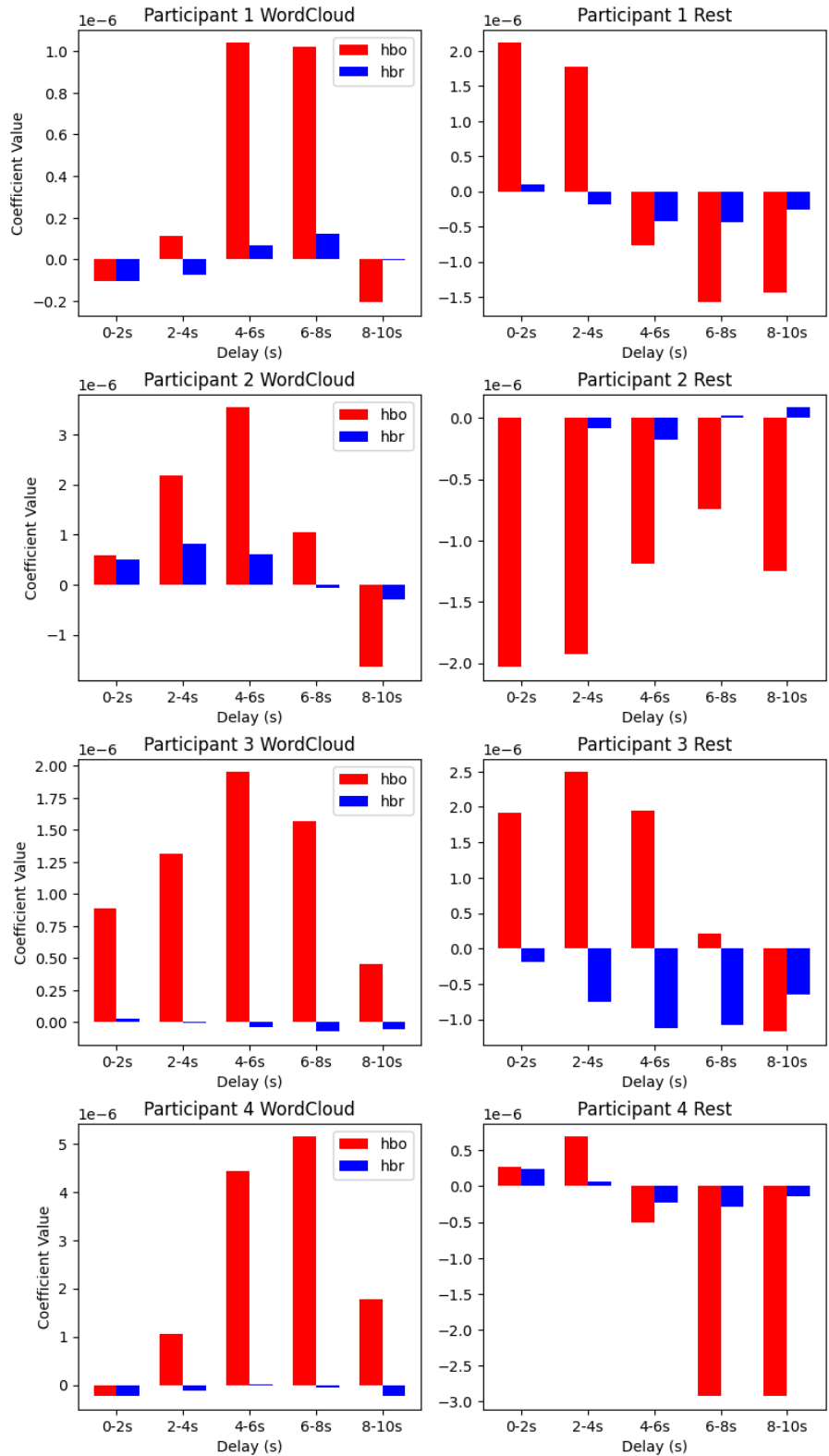

*Figure 4. FIR model coefficient values across delays showed the estimated shape of the BOLD response for each participant across word cloud and resting conditions.*

## 3.3 Imagined speech decoder: individual participant test results

Table 2 below shows test metrics for imagined speech decoders trained and tested with individual data. The metrics that measure both exact and semantic similarities between the predicted continuation texts as hypotheses, and the ground truth continuation from typed imagined sentences as references. The table heading represents 4 different experimental conditions - 1) Context input only, where only context input text inputs are used for the LLM to generate continuation predictions; 2) Brain signal only, where brain signals without context input are used for continuation predictions; 3) Brain+context input, where brain signals are converted to LLM embeddings and concatenated to context input embeddings, before being inputted to LLM for continuation predictions (the prompt tuning approach); 4) Brain signal Permutation+context only, where brain signals from another permuted trial are paired with context input from the current trial for continuation generation. This condition tests specifically whether the prompt tuning approach with unrelated but distribution conforming brain data improves the quality of generated continuations, acting as a stringent control condition to 3). The table shows both metric values and the corresponding t-test p-values between brain and permutation inputs (experiments 3 and 4).The metrics that are higher for brain compared to permutation inputs in values are highlighted in blue and those that are statistically significant (t-test $p<0.05$) are highlighted in green and provided in bold.

Using either only the context inputs or the brain signals produced generally low test metrics (experiments 1 and 2 in Table 2), however, by combining context inputs with corresponding brain signals (experiment 3) or permuted brain signals (experiment 4), there are sizable improvements across all test metrics (when comparing experiment 1 and 2 with 3 and 4 most t-test p-values $< 0.05$, except for values marked with † indicating $p>0.05$ for comparison with experiment 3, or with ‡ for comparison with experiment 4 in Table 2). Notably, when comparing generated continuation from brain and permutation inputs with the ground truth continuation (experiments 3 and 4 in Table 2), the exact match metric (BLEU-1) are statistically significantly higher for brain compared to permutation input for 3 out of 4 participants, and the semantic similarity metric (BERT P) are statistically significantly higher for 2 out of 4 participants with brain inputs, with an additional participant approaching significance. For participant 1, BERT P ($p = 0.02$) and BLEU-1 ($p = 0.004$) showed significant improvements (see definition in section 2.5.2 Data preparation). For participant 2, BLEU-1 ($p < 0.001$) and METEOR ($p = 0.017$) demonstrated significant enhancements, while BERT P ($p = 0.09$) approached significance, indicating a trend towards improved performance. For participant 4, BERT F1 ($p = 0.05$), BERT P ($p = 0.01$), and BLEU-1 ($p = 0.027$) were significantly better than the permutation. As for participant 3, the metrics BLEU-1, METEOR, ROUGE-L and WER are better than permutation but did not reach statistical significance.

Overall, these results highlight the capability of our prompt tuning-based continuous imagined speech decoder to leverage brain signals for generating text continuations that are both semantically and exactly similar to the ground truth. The consistent performance across multiple participants, as evidenced by significant improvements in key metrics, shows the robustness and potential of our approach in decoding brain signals for continuous speech decoding.

Table 2. Individually trained imagined speech decoders results.
Metrics measure exact and semantic similarity between predicted continuation against ground truth continuation texts for 200 test sentences. Numbers in brackets represent different experimental conditions. P-values are calculated with t-tests. A fixed delay of 6s is applied to all participants' data. (blue - brain input has outperformed the permutation input, green - brain input has outperformed the permutation input and is statistically significant with p<0.05. † indicating statistical insignificance p>0.05 for comparison with experiment 3, or with ‡ for comparison with experiment 4).)

| Participant | Test metrics | Context only vs ground truth (1) | Brain signal only vs ground truth (2) | Brain +context vs ground truth (3) | Permutation +context vs ground truth (4) | Brain (3) vs permutation (4) p-values |
|---|---|---|---|---|---|---|
| Participant 1 | BERT F1 | 0.865 | 0.857 | 0.878 | 0.875 | 0.173 |
| | BERT P | 0.848 | 0.843 | **0.869** | **0.862** | *0.02 |
| | BERT R | 0.883 | 0.872 | 0.889 | 0.889 | 0.877 |
| | BLEU-1 | 0.179 | 0.097 | **0.265** | **0.224** | *0.004 |
| | METEOR | 0.13 | 0.056 | 0.185 | 0.166 | 0.153 |
| | ROUGE L | ‡0.178 | 0.094 | 0.215 | 0.204 | 0.485 |
| | WER | 0.896 | 0.973 | 0.839 | 0.849 | 0.518 |
| Participant 2 | BERT F1 | 0.865 | 0.861 | 0.881 | 0.878 | 0.32 |
| | BERT P | 0.852 | 0.85 | **0.875** | **0.87** | 0.09 |
| | BERT R | 0.879 | 0.873 | 0.887 | 0.888 | 0.676 |
| | BLEU-1 | 0.17 | 0.12 | **0.308** | **0.253** | *<0.001 |
| | METEOR | 0.113 | 0.074 | **0.22** | **0.186** | *0.017 |
| | ROUGE L | 0.146 | 0.084 | 0.22 | 0.198 | 0.145 |
| | WER | 0.898 | 0.945 | 0.815 | 0.832 | 0.248 |
| Participant 3 | BERT F1 | 0.864 | 0.857 | 0.871 | 0.871 | 0.879 |
| | BERT P | 0.847 | 0.84 | 0.855 | 0.855 | 0.975 |
| | BERT R | 0.881 | 0.876 | 0.888 | 0.888 | 0.645 |
| | BLEU-1 | †‡0.166 | 0.073 | 0.191 | 0.189 | 0.319 |
| | METEOR | †‡0.116 | 0.046 | 0.138 | 0.136 | 0.403 |
| | ROUGE L | 0.16 | 0.065 | 0.214 | 0.212 | 0.319 |
| | WER | 0.904 | 0.968 | 0.873 | 0.875 | 0.319 |
| Participant 4 | BERT F1 | 0.861 | 0.859 | **0.869** | **0.866** | *0.05 |
| | BERT P | ‡0.853 | ‡0.855 | **0.862** | **0.857** | *0.01 |
| | BERT R | 0.87 | 0.863 | 0.876 | 0.875 | 0.496 |
| | BLEU-1 | 0.179 | 0.142 | **0.23** | **0.206** | *0.027 |
| | METEOR | 0.124 | 0.077 | 0.155 | 0.146 | 0.344 |
| | ROUGE L | 0.146 | 0.086 | 0.201 | 0.206 | 0.662 |
| | WER | 0.895 | 0.927 | 0.843 | 0.853 | 0.354 |

Table 3 provides example decoder-generated continuation sentences using both brain and permutation inputs (experiments 3 and 4), along with the context input and ground truth continuation texts. The context input represents the initial part of the imagined sentence. The LLM is then guided by either the brain inputs or the permutation inputs (which consists of brain data from another sentence) to complete the prediction. These examples are chosen because the brain predictions had higher BLEU-1 scores than the permutation predictions. Please note that these examples therefore represent the best examples of successful prompt tuning decoding.

The table illustrates that continuations generated from true brain data are closer to the ground truth (when metrics are statistically significant) in terms of both precise wording and semantic meaning. This trend is evident for participants 1, 2, and 4. In contrast, participant 3 had significantly fewer such sentences, consistent with their test metrics, which showed similar performance between brain and permutation predictions. This demonstrates the effectiveness of using brain data for generating text continuations closer to the actual imagined contents.

*Table 3. Predicted continuation examples from individually trained imagined speech decoders described above. The examples are from test cases that are held out in training.*

| Participant | Context input | Ground truth continuation | Brain input predicted continuation | Permutation input predicted continuation |
|---|---|---|---|---|
| Participant 1 | I used to frequently borrow | books from the library. | books from the library. | my dad's car when |
| | What is your main | desire in life. | goal in life. | hobby?. What |
| | A belief in a | higher power is faith. | higher power is often referred | religion is a set of |
| | Bananas grow in | warm tropical climates. | a tropical climate. | bunches of banana trees |
| | After the mile run | I felt refreshed. | I felt exhausted. | , I took a |
| | Next use this to | increase the efficiency. | create a link between | go to the next |
| Participant 2 | The street magic was | amazing to watch. | amazing to watch. | amazing and I could |
| | The horses tail was | long and bushy. | long and curly. | wagging back and forth |
| | Everything can be explored if | you can find the time. | you have the time. Everything | you know where to look. |
| | The phone connects older | people to their families. | people to their family and | phones to the internet. |
| | It is possible that technology will | change the world in the future. | improve in the future. It is | be the downfall of humanity in the |
| | The law judges whether something | is legal or not. | is a crime or not | is right or wrong. |

| Participant 3 | I tug what | I want. | I want to | I am doing |
|---|---|---|---|---|
| | What number is | your performance. | your favourite? | your favourite number |
| Participant 4 | Order and command are very | important in the military. | important in the military. | important in the world of |
| | The refrigerator must be maintained | at a certain temperature. | at a certain temperature. | to keep the food fresh |
| | Goals are not the same for | anyone and everyone, they are unique. | everyone, they are unique to each person | everyone and that's ok. Goals are |
| | I like to taste a variety | of flavours when i eat icecream. | of flavours in my food. I | of different foods and cuisines. I |
| | I do end up dreaming most | nights i just cant always remember. | nights before i go to sleep. | of the time when i go to |
| | If you can depend on someone there | is a high degree of reliance. | is a lot of trust involved. | is a lot to be said for |

## 3.4 Imagined speech decoder: multi-participant alignment and fine-tuning results

Table 4 presents the test metrics and statistical significance for the multi-participant alignment and fine-tuning experiment. To address time constraints that prevent participants from covering all topics, combining data from multiple participants can enhance the quantity and diversity of training topics. A simple ridge regression linear model aligns each participant's data to a unified latent feature space shared by all training participants, which is then used as input to the brain decoding model and training procedures described previously. Fine-tuning was conducted with 100 trials from the held-out test participant, followed by test metrics calculation using another 200 left out test trials. The same 4 experimental conditions were used as described in the previous section.

The results indicate that participants 2 and 3 achieved significantly higher BERT metrics (precision, recall, and F1 score) when using real brain signals over permutation inputs (experiments 3 and 4), while other metrics were higher but did not reach statistical significance. Participant 3, who did not achieve statistically significant performance when training an individual decoder with their own data, showed improved results when using the model trained with data from other participants. This suggests that combining data from multiple participants potentially enhances the model's ability to generalise across semantic meanings. However, participants 1 and 4 did not exhibit any statistically significant results comparing brain against permutation inputs. Given that participants 1 and 4 had distinct imagined sentence statistics (participant 1 used the most unique words per sentence, and participant 4 had the longest average imagined sentence length) (see *section 3.1* for dataset statistics per participant), it is possible that the fine-tuning process could not fully mitigate these variations. For completeness, context input only and brain signals only conditions were also tested (experiments 1 and 2), which showed lower test metrics compared to combined inputs, consistent with results from individually trained decoders. Overall, while multi-participant alignment training combined with fine-tuning shows potential for building a more versatile model, it may still be influenced by individual differences.

*Table 4. Multi-participant aligned decoder results. The decoder was trained on all other participants' data, and then fine tuned and tested on the test participant's data. Metrics measure exact and semantic similarity between predicted continuation against ground truth continuation texts for 200 test sentences. Numbers in brackets represent different experimental conditions. P-values are calculated with t-tests. A fixed delay of 6s is applied to all participants' data. (blue - brain input has outperformed the permutation input, green - brain input has outperformed the permutation input and is statistically significant with $p<0.05$ † indicating statistical insignificance $p>0.05$ for comparison with experiment 3, or with ‡ for comparison with experiment 4))*

| Test participant | Test metrics | LLM +context vs ground truth (1) | Brain signal only vs ground truth (2) | Brain +context vs ground truth (3) | Permutation +context vs ground truth (4) | Brain (3) vs permutation (4) p-values |
|---|---|---|---|---|---|---|
| Participant 1 | BERT F1 | 0.867 | 0.864 | 0.887 | 0.886 | 0.562 |
| | BERT P | 0.851 | 0.852 | 0.876 | 0.875 | 0.749 |
| | BERT R | 0.884 | 0.876 | 0.9 | 0.898 | 0.375 |
| | BLEU-1 | 0.194 | 0.118 | 0.299 | 0.289 | 0.549 |
| | METEOR | 0.14 | 0.079 | 0.235 | 0.228 | 0.595 |
| | ROUGE L | 0.193 | 0.106 | 0.288 | 0.28 | 0.614 |
| | WER | 0.866 | 0.939 | 0.776 | 0.775 | 0.97 |
| Participant 2 | BERT F1 | 0.864 | 0.865 | **0.879** | **0.873** | **\*0.02** |
| | BERT P | 0.85 | 0.857 | **0.869** | **0.863** | **\*0.045** |
| | BERT R | 0.879 | ‡0.873 | **0.889** | **0.884** | **\*0.011** |
| | BLEU-1 | 0.172 | 0.152 | 0.252 | 0.227 | 0.107 |
| | METEOR | 0.126 | ‡0.098 | 0.184 | 0.164 | 0.161 |
| | ROUGE L | 0.147 | ‡0.092 | **0.207** | **0.168** | **\*0.018** |
| | WER | 0.888 | 0.921 | 0.822 | 0.847 | 0.111 |
| Participant 3 | BERT F1 | 0.858 | 0.863 | **0.878** | **0.87** | **\*0.001** |
| | BERT P | ‡0.845 | 0.854 | **0.869** | **0.857** | **\*<0.001** |
| | BERT R | 0.873 | 0.873 | **0.888** | **0.884** | **\*0.047** |
| | BLEU-1 | 0.148 | 0.134 | 0.249 | 0.234 | 0.32 |
| | METEOR | 0.095 | 0.083 | 0.181 | 0.169 | 0.352 |
| | ROUGE L | 0.122 | 0.082 | 0.208 | 0.204 | 0.82 |
| | WER | 0.928 | 0.93 | 0.824 | 0.844 | 0.17 |
| Participant 4 | BERT F1 | 0.86 | 0.855 | 0.874 | 0.873 | 0.76 |
| | BERT P | ‡0.851 | 0.847 | 0.869 | 0.867 | 0.331 |
| | BERT R | 0.869 | 0.863 | 0.879 | 0.88 | 0.521 |
| | BLEU-1 | 0.171 | 0.103 | 0.248 | 0.232 | 0.278 |
| | METEOR | 0.118 | 0.06 | 0.184 | 0.169 | 0.299 |
| | ROUGE L | 0.14 | 0.071 | 0.184 | 0.181 | 0.797 |

| | WER | 0.915 | 0.951 | 0.84 | 0.846 | 0.692 |

## 3.5 Imagined speech detection

Our XTC classification model showed above chance results in distinguishing brain signals from imagined speech from rest (Figure 5). A total of 386, 532, 706, and 710 trials were included for Participant 1, 2, 3, and 4, respectively. The total number of trials consisted of an equivalent number of imagined speech and rest trials (i.e., imagined speech trials = ½ of total trials, rest trials = ½ of total trials). Please note that the dataset was balanced to ensure the same number of imagined speech and rest trials. Overall, our XTC model achieved an average accuracy of ~76% (p < 0.001, chance: 50%) when considering averaged accuracies across folds for the 4 subjects included in this test (max accuracy across subjects: 78%, p < 0.001) . Our best participant (participant 2) reported a best average accuracy of ~88% across the 3 folds in cross validation (p < 0.001) and max accuracy of ~90% (p < 0.001). The classifier was found to perform worse on participants 1 and 4, although it still performed well above chance (p < 0.001). See *Supplementary Table S1* for detailed results.

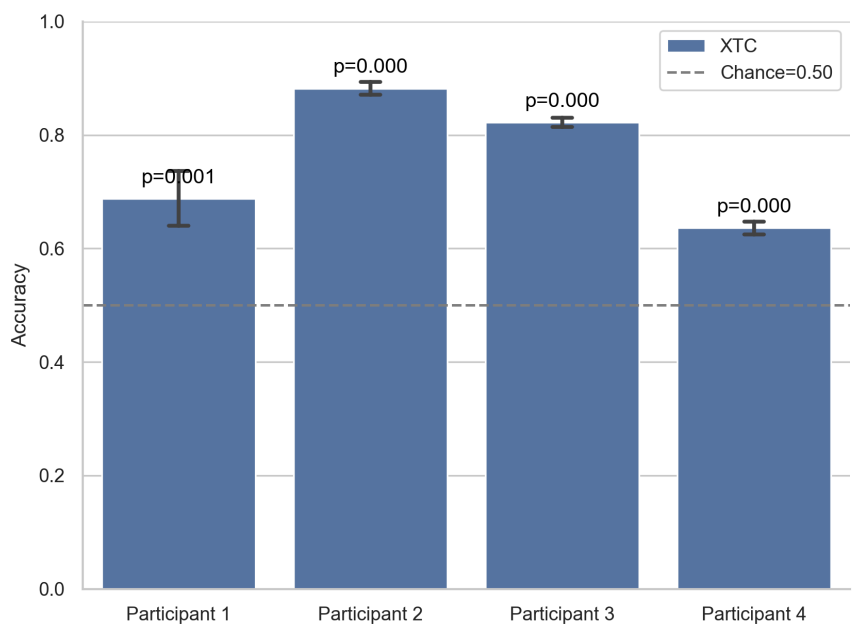

*Figure 5. Summary of XTC classifier model performance (accuracy %) in classifying imagined speech vs rest when using fully preprocessed fNIRS data. Data, standard deviation and p-values are reported for all participants. Comparison to chance (50%) is also reported.*

## 3.6 Brain areas underlying imagined speech

Figure 6 illustrates the surface cortex plots of HbO activations of the contrast 'word cloud > rest' for all four participants included in this study. Stimulus durations of 7s from each word cloud keyword highlight onset was considered. The colour bar represents the z-score values of the contrast. This analysis identified a recurrent pattern of multiple brain regions recruited during imagined speech, including the lateral temporal cortex, the dorsolateral prefrontal cortex (DLPFC) and the visual processing areas in the occipital region (Glasser et al., 2016). These regions are consistent with the previous research on covert/imagined speech (Petersen et al., 1988, Brumberg et al., 2016, Zhang et al., 2024), further proving that high density fNIRS is capable of capturing imagined speech processing in the brain.

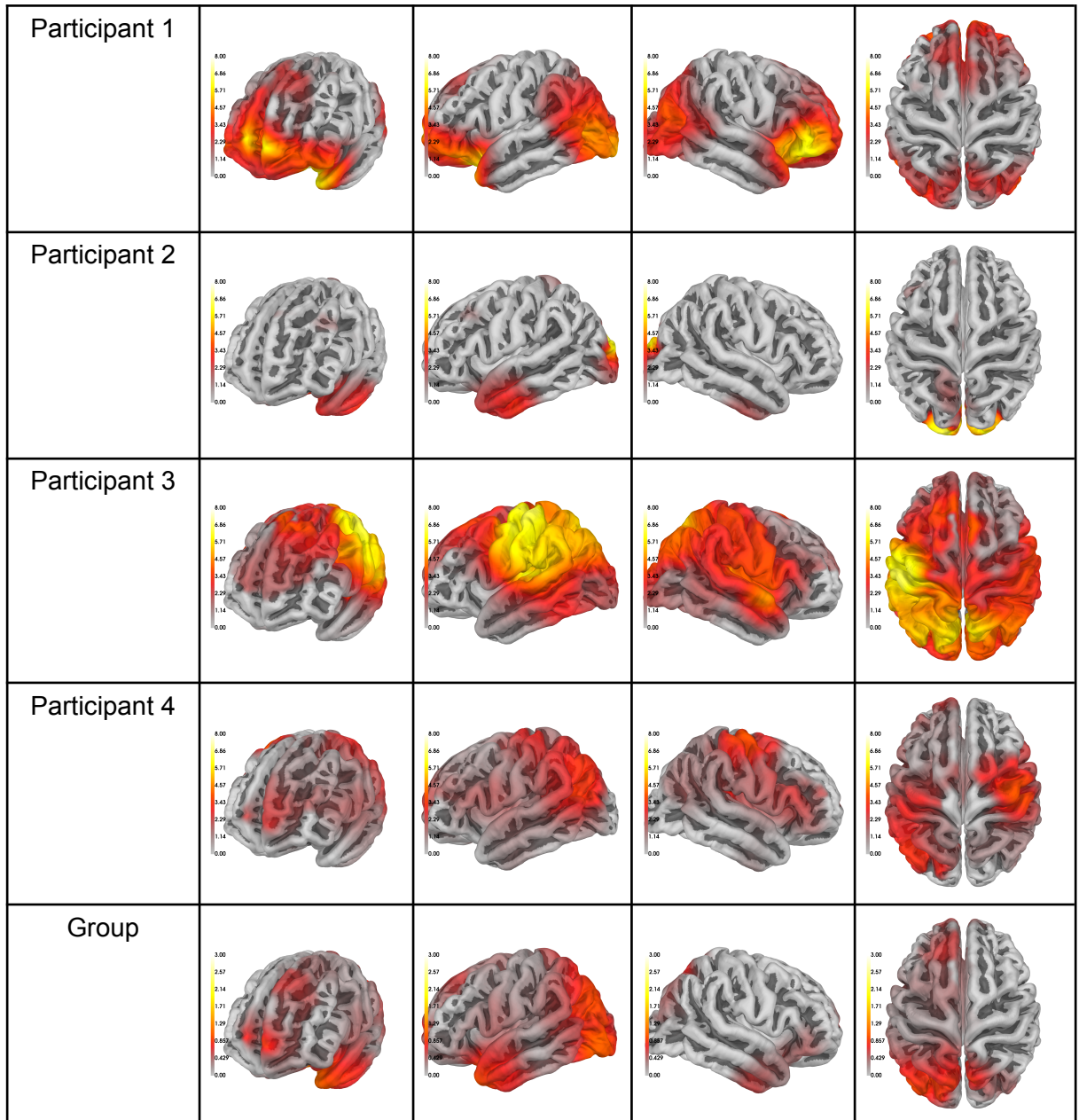

*Figure 6. Surface cortex plots showing HbO activations with the contrast word cloud > rest across all participants. The colorbar indicates z-score values.*

# 4. Discussion

We expanded our previous research MindGPT (Zhang et al., 2024) on the practicality of implementing a high-density fNIRS-based BCI system for deciphering imagined sentences as well as their semantic content. First, instead of classifying a limited number of sentences as in MindGPT, we now focused on developing an open vocabulary, continuous decoder for imagined speech fNIRS data. Our study introduced a novel paradigm for collecting high-density fNIRS data on imagined speech by prompting participants with topic words and keywords and having them imagine their own sentences. Ground truth sentences were obtained by having participants type them out on a computer using a keyboard after the imagined period. We then adapted the prompt tuning-based model from Ye et al., (2024) for generating continuous natural text from brain signals. Utilising the LLM Llama2-7b with a custom brain encoding model, brain signal guided inputs demonstrated improved language and semantic similarity in generated text continuation compared to permutation conditions. For individual training, the model with brain inputs achieved statistically significant higher BLEU-1 scores in 3 out of 4 participants as compared with permutation inputs, and significantly higher BERT P scores in 2 out of 4 participants (an additional one approaching significance). For multi-participant alignment, the model trained on multiple participants' data and fine tuned on the held out participant's data achieved statistically significant higher BERT F1/P/R scores in 2 out of 4 participants with brain compared to permutation inputs. Additionally, we were able to surpass our previous results on detecting imagined speech activity versus resting brain activity and achieved a 10% increase in decoder accuracy, and identified brain activations consistent with imagined speech research. Our findings suggest that using high-density fNIRS together with sophisticated artificial intelligence methods like LLM and prompt tuning could elevate prospects for the development of effective human-AI interaction, an interesting potential unveiled through this work.

Imagined speech paradigms in previous studies have predominantly focused on single words (Naseer and Hong, 2014), concept categories (Cao et al., 2018; Rybář et al., 2021), or memorised paragraphs (Tang et al., 2023; Zhang et al., 2024). These designs may not be optimal for data collection aimed for developing an open-vocabulary continuous imagined speech decoder. The reasons include their limited vocabulary or the absence of a time-bound ground truth, especially when participants recite long paragraphs from memory. To address these limitations, we introduced the word cloud paradigm. This new approach prompts participants to imagine new sentences within a specific time window, allowing each ground truth sentence to be accurately aligned with the corresponding section of brain data. The topic-based blocks enabled participants to easily recall the imagined sentences after the imagination period concluded. Additionally, the recall and typing activities had minimal impact on brain activity during the imagination phase. Crucially, this design eliminates the confounding effects of memory recall or auditory processing when using a metronome for pacing during the imagination phase, which can affect brain activity related to imagined speech and hinder decoder performance. Moreover, the word cloud paradigm is easily expandable with additional topics of varying semantic meanings and requires minimal training for new participants. Thus, it has potential as a novel and valuable addition to existing imagined speech paradigms.

For developing the continuous speech decoder, we adopted a different approach from the ridge regression-based methods used in previous studies (Pereira et al., 2018; Tang et al., 2023), by incorporating large language models (LLMs) into our decoder. Recent studies have increasingly leveraged LLMs to decode brain signals for various downstream tasks, particularly perceived speech and visual tasks, across different imaging modalities such as EEG (Duan et al., 2023) and fMRI (Tang et al., 2023; Xia et al., 2024; Ye et al., 2024). These studies have demonstrated the feasibility and effectiveness of utilising LLMs' extensive knowledge and comprehensive vocabulary in a generative setting for brain signal decoding.

Additionally, many studies (e.g., Caucheteux et al., 2022) have shown that language representations from the human brain and language models can be mapped to and from each other. Ye et al. (2024) further extended this research into a more brain-computer interface (BCI)-oriented approach, aligning brain recordings with language representations in LLMs and leveraging this alignment for language generation. Typically, language models generate coherent language based on contextualised representations extracted from text prompts. Advanced techniques such as prompt tuning (Liu et al., 2022) enable continuous time series data to be encoded into LLM embeddings and incorporated into the input context prompt, allowing the LLM to adapt to specific tasks, such as brain decoding, by training with a relatively small number of examples and without modifying its large number of parameters. By enriching these contextualised representations with information from brain recordings, it is possible for the LLM to generate language that more accurately reflects the semantics of these brain signals.

While adapting the prompt tuning model (Ye et al., 2024) for our continuous imagined speech decoder, we implemented several distinct improvements to optimise it for high-density fNIRS data. Firstly, we applied a finite impulse response (FIR) model to identify the appropriate BOLD delay specifically for the word cloud task. Unlike most fMRI decoding studies that use a fixed BOLD delay (e.g., Tang et al., 2023), this customization is crucial for fNIRS signals, which generally have a weaker signal-to-noise ratio (SNR) compared to fMRI. Offsets in delays can significantly impact decoding performance (Cui et al., 2011). Our results showed that the FIR model derived BOLD delays differed from conventional values (e.g., 8s used in Tang et al., 2023 and Ye et al., 2024) and produced better results. Secondly, we modified the architecture of the brain encoding model. We utilised a Seq2Seq model with transformers.. This approach leverages the self-attention mechanism of transformers and the contextual understanding of Seq2Seq models to dynamically model different time steps, capturing temporal dependencies more effectively (Sutskever et al., 2014; Vaswani et al., 2017). This adaptation is particularly beneficial for fNIRS data, which has a higher temporal resolution compared to fMRI data. Finally, we included a multi-participant alignment and fine-tuning experiment to demonstrate that combining data from multiple participants can potentially enhance the decoder's performance. Inspired by Scotti et al. (2024), we incorporated a simple ridge regression model to align multiple participants' brain data into a shared latent feature space before inputting it to the brain decoding model. This approach showed improvements in test metrics, particularly for participant 3, who did not show better results for brain inputs when training with their own data alone. In conclusion, our enhancements to the prompt tuning model—including more accurate BOLD delay estimation, Seq2Seq transformer architecture for the brain encoding model, and multi-participant data alignment—significantly improve continuous imagined speech decoding from high-density fNIRS data.

Both the individually-trained participant test results and the multi-participant alignment and fine-tuning outcomes highlight both the potential and the challenges of decoding continuous imagined speech from brain signals. For individually trained decoders, significant improvements in key test metrics such as BLEU-1 and BERT P for three out of four participants in individual training highlight the effectiveness of the prompt tuning-based text generation method. However, the results also reveal limitations, as participant 3, despite showing improvements for brain over permutation inputs in several test metrics, did not achieve statistical significance. Given that the same participant later achieved significantly higher BERT scores with brain over permutation inputs in the multi-participant alignment and fine tuning experiment, it is possible that more data and sentences are needed for decoder training in order to achieve robust performances. In regards to the other metrics that did not show statistically significant improvement for brain inputs, this may be due to the ineffectiveness of the metrics in evaluating short sentences, particularly for ROUGE and WER, as they focus solely on exact matches and are overly sensitive to minor deviations such as synonyms or slight rephrasings, particularly in abstractive summaries (Zhang et al., 2024). This is also evident in the multi-participant alignment and fine-tuning experiment, where brain inputs in participants 2 and 3 showed significantly higher BERT scores that indicate semantic similarity only, while participants 1 and 4 did not show statistically significant results, likely due to their unique sentence construction habits, which the fine-tuning process struggled to adapt to.

We observed that the use of permuted context input, in conjunction with brain signals as the baseline can introduce inconsistencies, as the brain signals from another sentence could be from a similar context as the current sentence and the decoder may fail to distinguish it from the real context input, paired with the brain signals. Therefore, we permuted the context inputs instead to produce a better test metric to evaluate context only or brain signal only conditions. While brain signals from resting periods could be used as a baseline instead, they may also be contaminated with random thoughts and are more susceptible to movement artefacts, making them less reliable compared to permutation inputs. The noisy nature of brain data can limit how much the decoder generalises to novel test inputs, for example, this is demonstrated by the lower test metrics from the brain signal only experimental condition, as compared to context only condition where only consistent noise-free text token information is used for continuation predictions. Overall, these decoder related findings illustrate the complexity of decoding continuous imagined speech from brain signals, where individual differences can significantly impact performance. Although multi-participant alignment and fine-tuning hold potential for creating more versatile decoders, their effectiveness is limited by variability in individual brain signal patterns, linguistic habits, and noise during data collection. Further research and refinement are needed to better understand and address these individual differences to enhance the robustness of brain signal decoding methods.

Finally, we employed an XTC model to differentiate between imagined speech and resting states, achieving an average accuracy of 76% across four subjects. These findings corroborate our previous study, MindGPT (Zhang et al., 2024), showing a 10% improvement in decoding accuracy between imagined speech and resting states in the current study, likely attributable to the increased engagement in imagined speech and enhanced data quality from the word cloud paradigm. However, it should be noted that the increased visual region activations from keyword highlighting may also contribute to the improved decoding accuracy. A previous study (Lesaja et al., 2022) showed a ~90% accuracy in detecting overt speech using invasive stereo-EEG. In contrast, our method shows the feasibility of decoding covert (imagined) speech using non-invasive neuroimaging (fNIRS), with a reasonably high level of accuracy. Furthermore, brain activation analysis additionally revealed significant brain activations in regions such as the dorsolateral prefrontal cortex (DLPFC), lateral temporal cortex, and visual areas when contrasting imagined speech with rest, supporting our interpretations and previous research on imagined speech decoding (Brumberg et al., 2016; Eggebrecht et al., 2014; Petersen et al., 1988).

# 5. Limitations and further work

## 5.1 Limitations

While our results in this paper show a breakthrough advancement in this field, it is crucial to acknowledge limitations such as the need for improved sensitivity in decoding, evident in the non-significantly higher BERT R scores in individually trained decoders with brain over permutation inputs. The results suggest that the current decoder is accurate in its generated continuation text but could not cover all the test cases. This can be improved by using more training data, either through multi-participant alignment or synthetic data augmentation. While multi-participant alignment tested in this study appears to improve BERT R scores, it also has its own limitations where only 2 participants achieved statistically significant results. The reliance on a limited sample size also restricts the generalisability of our findings, as a broader and more diverse participant pool could yield more robust and universally applicable results, particularly for training a multi-participant aligned model.

Additionally, our results highlighted the variability in individual responses to the decoding process and the need for personalised approaches. These limitations underscore the necessity for ongoing research and development in this topic, aimed at refining both the brain encoding model and the multi-participant alignment procedure, in order to enhance the decoder's ability to better adapt to individual participants.

Finally, it is evident from the test metrics in Table 2 that the context input (experiment 1) plays a key role in the decoding process. The use of brain signals only (experiment 2) is not sufficient to precisely decode the continuation. This can be partly due to the noisy nature of the brain signals, which makes the predicted LLM embeddings less accurate. Overall, it can be seen that the combination of context input and brain signals (experiment 3) offer the best performance, in most cases. This further suggests the need for an organic brain-computer interaction for devising the future of human-AI interaction.

## 5.2 Further work

The following steps can be taken to overcome the aforementioned limitations. Firstly, more data can be collected from a larger and more diverse pool of participants, as well as incorporating a wider range of semantic meanings for topics used in the word cloud paradigm, to validate our findings and improve decoder performance. Secondly, leveraging other advanced AI techniques can improve the current prompt-tuning based decoder. Specific brain signal features can be aligned with specific LLM representations, and other fusion methods between text and brain signal generated embeddings can be explored (Jin et al., 2024). Teaching LLMs to make use of other pre-trained brain signal based time series models will likely improve outcomes too. Additionally, brain signal processing steps can be further optimised, including signal preprocessing, noise removal and feature extraction. Given that fNIRS data has generally lower SNR compared to modalities such as fMRI, ensuring high-quality data input to the training pipeline is essential.

Building a real-time system that is able to continuously decode imagined speech from streamed brain data will be a future goal too. This current model exhibits characteristics that can be utilised for such applications, however, reducing LLM inference time and hardware usage, as well as building an efficient real-time data preprocessing pipeline will be essential for realising a real-time version of this model.

# 6. Conclusion

Our study advanced human-AI interaction by developing an open-vocabulary, continuous decoder for imagined speech using high-density fNIRS data. We introduced a novel word cloud paradigm, which improved data quality and allowed participants to generate imagined sentences with varied semantic meanings. We then applied a prompt tuning-based approach to leverage the Llama2 LLM for text generation guided by imagined speech brain signals. Significant improvements in key metrics, such as BLEU-1 and BERT P scores, were observed in three out of four participants, demonstrating the effectiveness of our approach. The multi-participant alignment and fine-tuning experiment further showed that combining data from multiple participants can enhance decoder performance, with statistically significant improvements observed in BERT scores for two out of four participants.

Overall, this work demonstrates the potential of using AI techniques to decode imagined speech for a direct communication with computers and AI with imagined speech, using a portable headgear.

# Supplementary Information

**Supplementary Table S1.** Decoding accuracy performance of the XTC model when inputting fully preprocessed data from participants 2, 3 and 4, assessed over 5 different seeds, and 3 k-folds. Best model accuracy and average model accuracy across the different folds and different seeds per participant are reported; as well as significance ($p < 0.05$). Averaged values across participants are also outlined.

| Supplementary Table S1. XTC Model tests results across all participants - fully preprocessed data | | | | | | |
|---|---|---|---|---|---|---|
| Model | Best Accuracy | Avg Accuracy | p-value | Distribution | | Seed |
| | | | | Imagined | Rest | |
| Participant 1 | 0.78 | 0.74 | < 0.001 | 0.60 | 0.61 | 0 |
| Participant 1 | 0.71 | 0.69 | < 0.001 | 0.70 | 0.67 | 6 |
| Participant 1 | 0.63 | 0.61 | < 0.001 | 0.74 | 0.71 | 12 |
| Participant 1 | 0.73 | 0.71 | < 0.001 | 0.69 | 0.62 | 24 |
| Participant 1 | 0.71 | 0.69 | < 0.001 | 0.70 | 0.61 | 42 |
| **Participant 1 Avg.** | **0.71** | **0.69** | **< 0.001** | **0.69** | **0.64** | **-** |
| Participant 2 | 0.88 | 0.87 | < 0.001 | 0.82 | 0.89 | 0 |
| Participant 2 | 0.88 | 0.88 | < 0.001 | 0.80 | 0.87 | 6 |
| Participant 2 | 0.89 | 0.88 | < 0.001 | 0.78 | 0.87 | 12 |
| Participant 2 | 0.94 | 0.90 | < 0.001 | 0.79 | 0.88 | 24 |
| Participant 2 | 0.89 | 0.88 | < 0.001 | 0.82 | 0.88 | 42 |
| **Participant 2 Avg.** | **0.90** | **0.88** | **< 0.001** | **0.80** | **0.88** | **-** |
| Participant 3 | 0.83 | 0.81 | < 0.001 | 0.81 | 0.83 | 0 |
| Participant 3 | 0.84 | 0.83 | < 0.001 | 0.79 | 0.83 | 6 |
| Participant 3 | 0.83 | 0.82 | < 0.001 | 0.80 | 0.83 | 12 |
| Participant 3 | 0.85 | 0.82 | < 0.001 | 0.78 | 0.81 | 24 |
| Participant 3 | 0.84 | 0.83 | < 0.001 | 0.79 | 0.77 | 42 |
| **Participant 3 Avg.** | **0.84** | **0.82** | **< 0.001** | **0.79** | **0.81** | **-** |
| Participant 4 | 0.67 | 0.65 | < 0.001 | 0.65 | 0.66 | 0 |
| Participant 4 | 0.64 | 0.64 | < 0.001 | 0.61 | 0.65 | 6 |
| Participant 4 | 0.63 | 0.62 | < 0.001 | 0.65 | 0.69 | 12 |
| Participant 4 | 0.65 | 0.64 | < 0.001 | 0.63 | 0.68 | 24 |
| Participant 4 | 0.64 | 0.63 | < 0.001 | 0.69 | 0.70 | 42 |
| **Participant 4 Avg.** | **0.65** | **0.64** | **< 0.001** | **0.65** | **0.68** | **-** |
| **Tot. Avg.** | **0.78** | **0.76** | **< 0.001** | **0.73** | **0.75** | **-** |

**Supplementary Figure S1.** Description of the finger tapping paradigm conducted at the start and end of each session in this study (A). Participants' brain activations in response to 10s of right or left finger tapping trials and their localisation within the cortex (B-C-D-E). Brain activation visualisations built using general linear modelling (GLM, Friston et al., 1994).

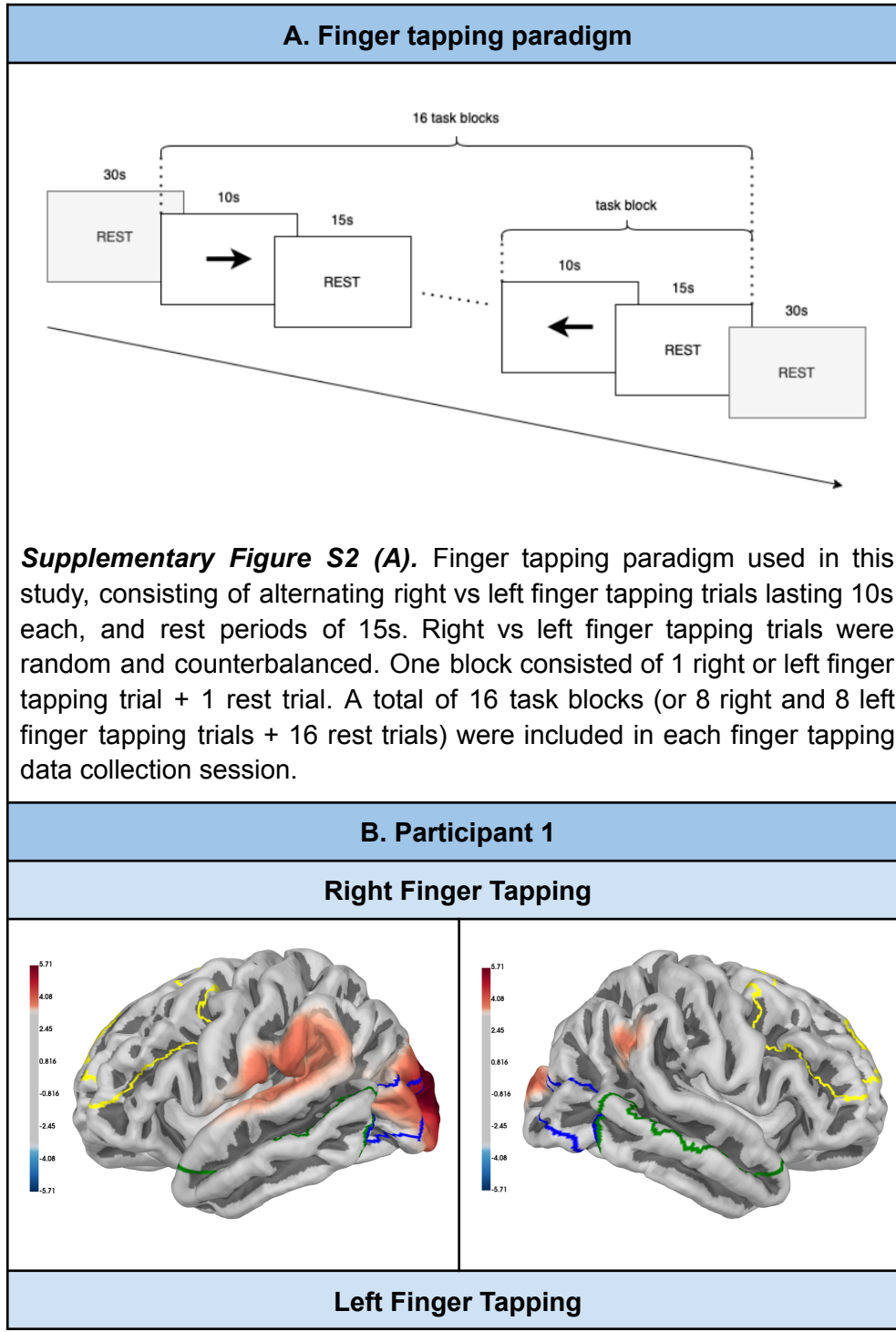

### A. Finger tapping paradigm

***Supplementary Figure S2 (A).*** Finger tapping paradigm used in this study, consisting of alternating right vs left finger tapping trials lasting 10s each, and rest periods of 15s. Right vs left finger tapping trials were random and counterbalanced. One block consisted of 1 right or left finger tapping trial + 1 rest trial. A total of 16 task blocks (or 8 right and 8 left finger tapping trials + 16 rest trials) were included in each finger tapping data collection session.

### B. Participant 1

#### Right Finger Tapping

#### Left Finger Tapping

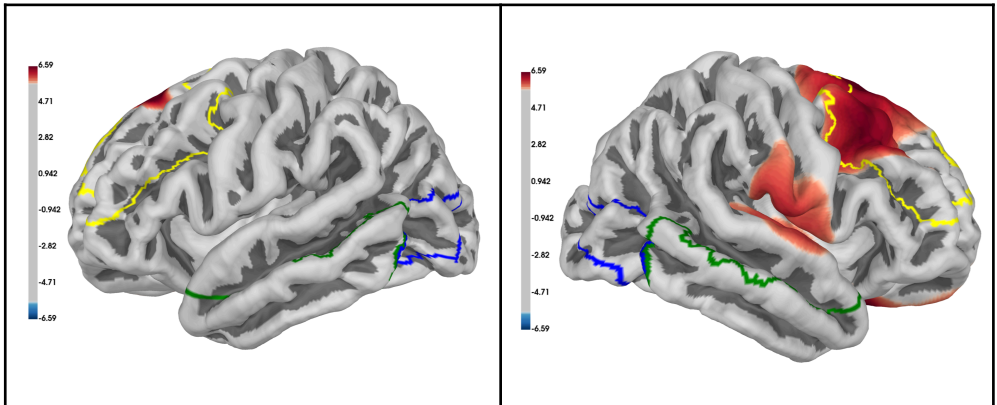

**C. Participant 2**

**Right Finger Tapping**

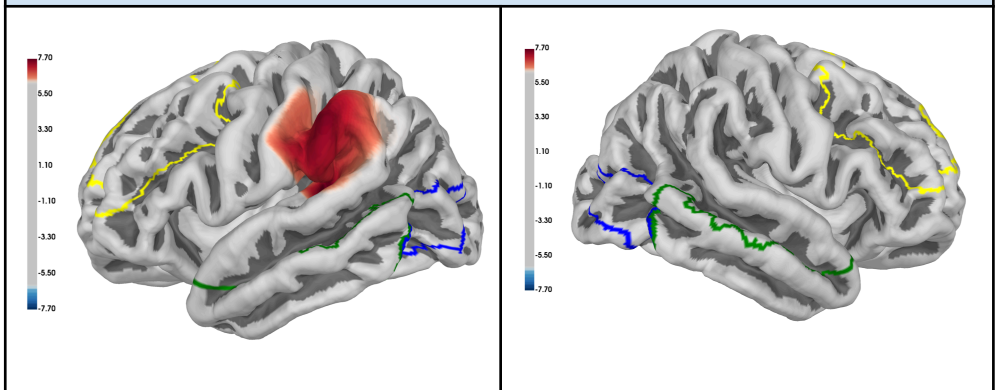

**Left Finger Tapping**

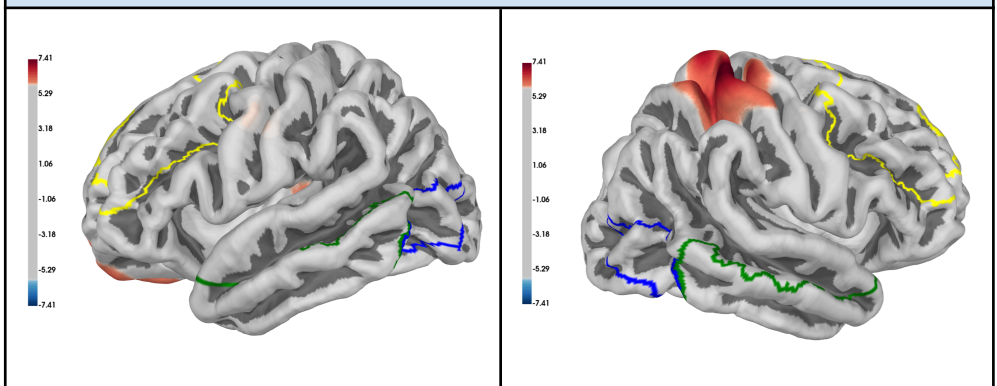

**D. Participant 3**

**Right Finger Tapping**

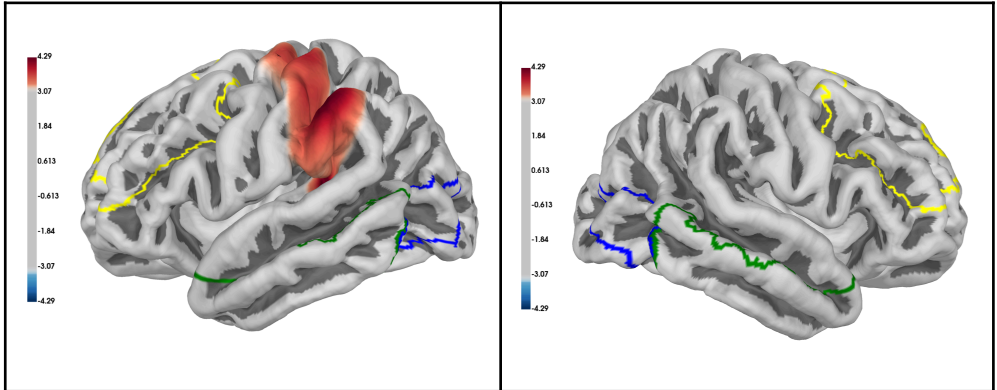

**Left Finger Tapping**

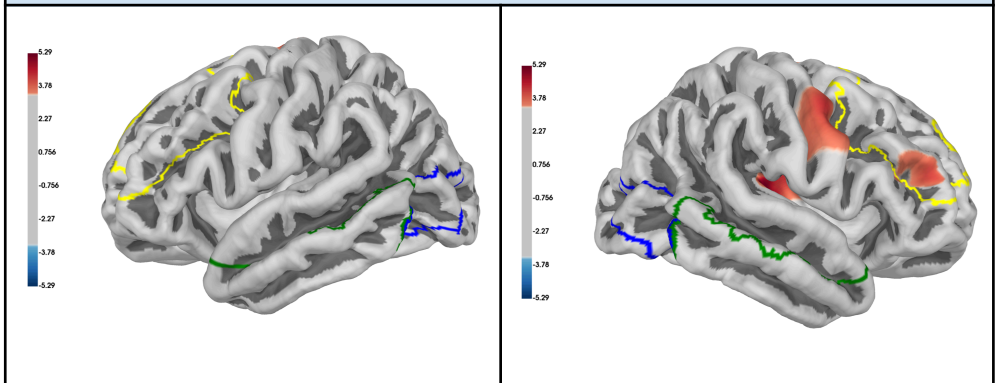

**E. Participant 4**

**Right Finger Tapping**

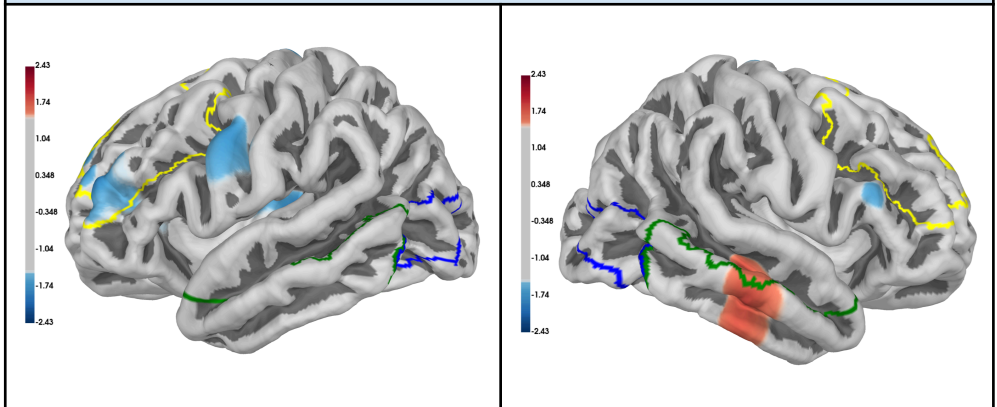

**Left Finger Tapping**

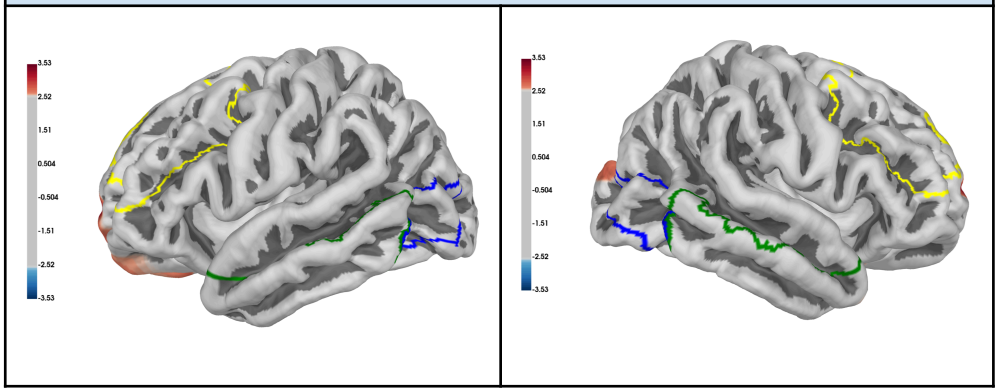

MindPortal